\documentclass[12pt]{article}
\usepackage{mathrsfs}
\usepackage{amssymb}
\usepackage{dsfont}
\normalsize
\input{epsf}

\begin{document}
\addtolength{\baselineskip}{.20mm}
\newlength{\extraspace}
\setlength{\extraspace}{2mm}
\newlength{\extraspaces}
\setlength{\extraspaces}{2mm}

\newcommand{\newsection}[1]{
\vspace{15mm} \pagebreak[3] \addtocounter{section}{1}
\setcounter{subsection}{0} \setcounter{footnote}{0}
\noindent {\Large\bf \thesection. #1} \nopagebreak
\medskip
\nopagebreak}
\newcommand{\newsubsection}[1]{
\vspace{1cm} \pagebreak[3] \addtocounter{subsection}{1}
\addcontentsline{toc}{subsection}{\protect
\numberline{\arabic{section}.\arabic{subsection}}{#1}}
\noindent{\large\bf %\thesection.
\thesubsection. #1} \nopagebreak \vspace{3mm} \nopagebreak}
\newcommand{\ba}{\begin{eqnarray}
\addtolength{\abovedisplayskip}{\extraspaces}
\addtolength{\belowdisplayskip}{\extraspaces}

\addtolength{\belowdisplayshortskip}{\extraspace}}

\newcommand{\be}{\begin{equation}
\addtolength{\abovedisplayskip}{\extraspaces}
\addtolength{\belowdisplayskip}{\extraspaces}
\addtolength{\abovedisplayshortskip}{\extraspace}
\addtolength{\belowdisplayshortskip}{\extraspace}}
\newcommand{\ee}{\end{equation}}
\newcommand{\STr}{{\rm STr}}
\newcommand{\figuur}[3]{
\begin{figure}[t]\begin{center}
\leavevmode\hbox{\epsfxsize=#2 \epsffile{#1.eps}}\\[3mm]
\parbox{15.5cm}{\small
\it #3}
\end{center}
\end{figure}}
\newcommand{\im}{{\rm Im}}
\newcommand{\calm}{{\cal M}}
\newcommand{\call}{{\cal L}}
\newcommand{\sect}[1]{\section{#1}}
\newcommand\hi{{\rm i}}
\def\bea{\begin{eqnarray}}
\def\eea{\end{eqnarray}}

\begin{titlepage}
\begin{center}

\vspace{3.5cm}

{\Large \bf{A New Type of Dark Energy Model}}\\[1.5cm]

{Yi Zhang $^{a,b,}$\footnote{Email: zhangyia@cqupt.edu.cn},}{Hui Li
$^{c,}$\footnote{Email:lihui@ytu.edu.cn },} \vspace*{0.5cm}

{\it} $^{a}$Department of Astronomy, Beijing Normal university,
\\  Beijing 100875, China

$^{b}$College Mathematics and Physics, Chongqing Universe of Posts and Telecommunications, \\ Chongqing 400065, China

$^{c}$ Department of Physics, Yantai University, Yantai 264005,
China

\date{\today}
\vspace{3.5cm}

\textbf{Abstract} \vspace{5mm}

\end{center}
In this paper, we propose a  general form of the equation of state
(EoS) which is the function of the fractional dark energy density
$\Omega_{d}$. At least, five related models, the cosmological
constant model,  the holographic dark energy model, the agegraphic
dark energy model, the modified holographic dark energy model and
the Ricci scalar holographic dark energy model are included in this
form. Furthermore, if we consider proper interactions, the
interactive variants of those models can be included as well. The
phase-space analysis shows that the scaling solutions may exist both
in the non-interacting and interacting cases. And the stability
analysis of the system could give out the attractor solution which
could alleviate the coincidence problem.
\end{titlepage}

%============================= section 1 ===================================
\section{Introduction}\label{sec1}
Nowadays there is a wide consensus among observational cosmologists
that our universe is accelerating
\cite{supernova,wmap,sdss,spergel}. However,  the essence of  this
acceleration is still an open question \cite{dark energy}. The
cosmological constant is the simplest explanation, but it has the
fine-tuning problem  which requires the observed cosmological
constant much smaller than the fundamental Planck scale, and the
coincidence problem as well: why the cosmological constant and the
matter have comparable energy density today even their evolution
behavior is so different.

Dynamical dark energy (DE)
 models have been proposed as alternatives to the cosmological
 constant, such as quintessence and phantom \cite{coinpro1,coinpro2}. In
principle, the dark energy problem may be one part of the puzzles of
quantum gravity. However, we have not invented a complete theory of
quantum gravity as yet. For all that, important progress in the
study of the black hole theory and string theory is the holographic
principle \cite{Cohen:1998zx}, which could be considered as a
fundamental principle of quantum gravity and then shed some light on
the DE problem. Its typical application in cosmology is holographic
dark energy \cite{Hsu,Li}, the agegraphic dark energy \cite{Cai},
modified holographic dark energy
 \cite{Gong} and Ricci scalar
dark energy \cite{Gao} as well. They are  dynamical models in which
the dark energy equation of state (EoS) $\omega_{d}$ can be written
as a function of the fractional dark energy density $\Omega_{d}$.

However,  we emphasize that such flexibility and generality are
particularly important to our research on $\omega_{d}$, not only
because they increase the range of possibility to be tested, but
also because in principle they may reduce the possibility of
misleading results an incorrect EoS parameterization can produce.
For detailed research
 on $\omega_{d}$, we assume a kind of holographic model in which
  the dark energy equation of state (EoS) $\omega_{d}$ can be written
as a function of the fractional dark energy density $\Omega_{d}$. If
we introduce a barotropic fluid to the system, the dynamics makes
the system an autonomous system.

In the previous work, many  studies focus on the stability property
of cosmological scaling solutions in an expanding universe in the
scalar field dark energy model. A phase-space analysis of the
spatially flat FRW (Friedmann-Robertson-Walker) models shows that
there exist cosmological scaling solutions which are the unique
late-time attractors \cite{d1,d2,d3,d4}. With the assumption of
$\omega_{d}$ being a function of $\Omega_{d}$,  it may be a
self-similar system that we can do some research on its phase space
from the point of stability analysis. In addition, the cosmological
evolution of the dark energy interacting with background perfect
fluid could be investigated  at the same time. We consider the cases
of dark energy interacting with background perfect fluid, while the
interaction terms are taken to be three different forms which are
familiar in the literature \cite{int1,int2,int3}. And the physical
consequence of these results should be given out as well. In this
paper, we also do many observational constraint discussions on
these interacting models \cite{refADE,refHDE,refRDE}.

This paper is organized as follows. In section \ref{sec2}, we
propose our dark energy model and present the physical background
that we refer to.
 In section  \ref{sec3}, we extend  to the interaction cases.
In section  \ref{sec4}, we  give out the results of  exact
phase-analysis. In section  \ref{sec5}, we try to analyze the four
examples in detail.
 Finally, a short summary will be presented.

\section{Physical background}\label{sec2}
The standard cosmology suggests that our universe goes through the
radiation dominated period, the matter dominated period, and now the
dark energy dominated period. For consistence, any theoretical
models should coincide with this history of the universe.

In this work, we assume that the geometry of space-time is described
by the flat FRW  metric which seems to be consistent with today's
cosmological observations
\begin{eqnarray}
 ds^{2}=-dt^{2}+a^{2}(t)\sum^{3}_{i=1}(dx^{i})^{2},
\end{eqnarray}
where $a$ is the scale factor. Based on the subject we are
interested in, we assume that there are two main components in the
universe: the background matter and the dark energy. The background
matter is assumed to be described by a perfect fluid with barotropic
equation of state
 \be
\label{baro}
 \rho_{m}'+3\gamma\rho_{m}=0,
 \ee
 where $\rho_{m}$ is the energy density of the background matter, a prime denotes the derivative with respect to the e-folding
time $N\equiv\ln a$, and the barotropic index $\gamma$ is
 a constant and satisfies $0<\gamma\leq2$. In particular, $\gamma=1$ and $\gamma=4/3$ correspond to dust matter and radiation, respectively.
And the dark energy component  leads to
\begin{eqnarray}
\label{Lambda1}
&& \rho_{d}'+3(1+\omega_{d})\rho_{d}=0.
\end{eqnarray}
where the  $\rho_{d}$ is the energy density of dark energy, and the
dark energy EoS parameter  is $\omega_{d}=p_{d}/\rho_{d}$.

With respect to the whole system, the Friedmann and  Raychaudhuri
equations read
\begin{eqnarray}
\label{FE}
&& H^{2}=\frac{\rho_{tot}}{3m_{pl}^{2}}=\frac{1}{3m_{pl}^{2}}(\rho_{m}+\rho_{d}),\\
 \label{RE}
&&\dot{H}=-\frac{1}{2m_{pl}^{2}}(\rho_{tot}+p_{tot})=-\frac{(1+\omega_{tot})\rho_{tot}}{2m_{pl}^{2}},
\end{eqnarray}
where $m_{pl}$ is the Planck mass and the index ``tot" notes the
variables for the whole system.

It is convenient to introduce the fractional energy densities
  \be
  \label{defineo}
 \Omega_{i}\equiv\rho_{i}/(3m_{pl}^{2}H^{2}),
  \ee
  with i being m or d. The Friedmann Equation can be rewritten as
 \be
 \label{constraint}
 \Omega_{m}+\Omega_{d}=1.
 \ee
 Furthermore, the EoS parameter can be reexpressed as
 \be
 \label{wt}
\omega_{tot}=\frac{\frac{\Omega_{m}}{\Omega_{d}}\omega_{m}+\omega_{d}}{1+\frac{\Omega_{m}}{\Omega_{d}}}.
 \ee

As the universe evolves, combining Eqs. (\ref{baro}),
(\ref{Lambda1}), (\ref{RE}) and (\ref{defineo}), we can get the
equations of motion for $\Omega_{i}$ separately
 \begin{eqnarray}
&& \Omega_{m}'= 3f_{n}\Omega_{m}\Omega_{d},\\
&& \Omega_{d}'= -3f_{n}\Omega_{d}\Omega_{m},
 \end{eqnarray}
with $f_{n}=\omega_{d}-(\gamma-1)$, where the index $n$ means there
is no interaction between the background matter and the dark energy.

However, we do not really have the exact form of $\omega_{d}$, so we
could not know the form of $f_{n}$. There is a kind of holographic
dark energy type model where
 $\omega_{d}$ is a function of
$\Omega_{d}$, which can make the dynamic system an autonomous
system. This kind of model has been widely researched.
 It was
firstly motivated from  the effective quantum field theory. Cohen,
et al. suggested that the quantum zero-point of a system with the
size $L$ should not exceed the mass of a black hole with the same
size,i.e. $L^{3}\rho_{d}<Lm_{pl}^{3}$, where $\rho_{d}$ is the
quantum zero-point energy density which we could use as dark energy.
Thus, the ultraviolet (UV) cutoff scale of a system is connected to
its infrared (IR) cut-off scale. Applying this idea to the whole
universe, the vacuum energy can be considered as DE. Choosing the
largest IR cutoff $L$ which saturates the inequality, we obtain the
holographic DE density
 \begin{eqnarray}\label{L}
 \rho_{d}=3c^{2}m_{pl}^{2}L^{-2},
  \end{eqnarray}
 where $c$ is an unknown constant
due to the theoretical uncertainties and can only be determined by
observations.

We can see that, with  $L=constant$, it recovers the cosmological
constant model. Interestingly, taking $L$ to be the size of the
current universe which is the Hubble radius $H^{-1}$, it yields a
wrong equation of the state for DE, but a correct  DE density which
is close to the observed values. Choosing different $L$'s, we can
get  different EoS which are listed in Tab.1. In particular,
holographic dark energy chooses the event horizon as the cutoff,
agegraphic dark energy chooses the age of the universe as the
cutoff, modified holographic dark energy model  relates the cutoff
to the blackhole mass, and Ricci scalar holographic model chooses
the value of Ricci scalar as cutoff. In each of the models, the dark
energy state of equation $\omega_{d}$ is related to the dark energy
fractional energy $\Omega_{d}$, which could reduce a self-similar
system.
%==================== table 1 ====================
 \begin{table*}[t]
\begin{center}
\label{tab1}
 \caption[crit]{The cutoff and the equation of state in cosmological
constant model, holographic dark energy model, agegraphic dark
energy model, modified holographic dark energy model and Ricci
scalar dark energy model. $c$, $n$, $\alpha$ and $\beta$ are all
related parameters in these models.}
\begin{tabular}{|c|c|c|c|}

    \hline\hline
Dark energy models&$L$&$\omega_{d}$ \\
\hline
Cosmological constant&Constant&$\omega_{d}=-1$ \\
Holographic&Event horizon&$\omega_{d}=-\frac{1}{3}(1+\frac{2\sqrt{\Omega_{d}}}{c})$ \\
 Agegraphic&Age of the universe&$\omega_{d}=-\frac{1}{3}(3-\frac{2\sqrt{\Omega_{d}}}{n})$ \\
   Modified holographic& Blackhole mass&$\omega_{d}=\frac{\alpha-2}{2-\alpha\Omega_{d}}$ \\
  Ricci &Ricci scalar&$\omega_{d}=-\frac{2}{3\beta^{2}}+\frac{1}{3\Omega_{d}}-\frac{\Omega_{m}}{\Omega_{d}}(\gamma-1)$ \\

\hline
 \hline
 \end{tabular}
\end{center}
\end{table*}
 For a
complete and exact examination, we could treat  $f_{n}$ as a general
form of the $\Omega_{d}$ which is
 \be
f_{n}(\Omega_{d})=\omega_{d}-w_{m}=\omega_{d}(\Omega_{d})-(\gamma-1),
 \ee
 where the form of $f_{n}$ or $\omega_{d}$ is determined by the unknown underlying theory.
Assuming $f_{n}$ exists, the system may be identified to be an
autonomous system. Then we can discuss the properties based on the
existence and the stability of its critical points.

\section{Interacting Effects}\label{sec3}
 Interacting models have been discussed largely
in order to  understand or alleviate the coincidence problem . The
basic idea is to consider the possible interaction between dark
energy and dark matter owing to the unknown nature of dark energy
and dark matter.
  In
this section, by including the interaction between the dark energy
and the background fluid, we will consider the change of the form
$f_{n}$. Although the interaction can significantly change the
cosmological evolution, with proper interactions, the system is
still an autonomous system.

Assuming the dark energy and the background matter exchange energy
through interaction term $Q$, the continuity equations become
\begin{eqnarray}
&& \rho_{d}'+3(1+\omega_{d})\rho_{d}=-Q,\\
 &&\rho_{m}'+3\gamma\rho_{m}=Q,
\end{eqnarray}
which still preserve the total energy conservation equation
$\dot{\rho}_{tot}+3H(\rho_{tot}+p_{tot})=0$.  When $Q=0$, there is
no interaction between matter and dark energy. When $Q>0$, the dark
energy will lose energy to the the background matter. When $Q<0$,
the dark energy will get energy from  the background matter. The
interaction term $Q$ can be assumed to be some special forms. For
convenience, we consider the following specific interaction forms:
\begin{eqnarray}
\label{case1}
& {\rm Case~(I)} &Q_{1}=3\gamma_{m}\rho_{m},\\
\label{case2}
  &{\rm Case~(II)}& Q_{2}=3\gamma_{d}\rho_{d},\\
  \label{case3}
  & {\rm Case~(III)}& Q_{3}=3\gamma_{tot}\rho_{tot}.
   \end{eqnarray}
We use the  indices $1,2,3$ notify the different interacting cases.
 We can
write  the effective EoS parameters for both  dark energy and
background matter
 \begin{eqnarray}
&&\omega_{de1}=\omega_{d1}(\Omega_{d})+\gamma_{m}\frac{1-\Omega_{d}}{\Omega_{d}},\,\,\,\,\omega_{me1}=\gamma-1-\gamma_{m}\\
&&\omega_{de2}=\omega_{d1}(\Omega_{d})+\gamma_{d},\,\,\omega_{me2}=\gamma-1-\gamma_{d}\frac{\Omega_{d}}{1-\Omega_{d}}\\
&&\omega_{de3}=\omega_{d1}(\Omega_{d})+\gamma_{tot}\frac{1}{\Omega_{d}},\,\,\omega_{me3}=\gamma-1-\frac{\gamma_{tot}}{1-\Omega_{d}}
 \end{eqnarray}
where the index $e$ means the  EoS parameter of dark energy or
matter is
 `` effective". Furthermore, we find
 \be
 \label{fj}
 f_{j}=\omega_{dej}-\omega_{mej}
 \ee
  where $j=1,2,3$, and
\begin{eqnarray}
&&f_{1}=f_{n}+\frac{\gamma_{m}}{\Omega_{d}}=\omega_{d}(\Omega_{d})-(\gamma-1)+\frac{\gamma_{m}}{\Omega_{d}},\\
&& f_{2}=f_{n}+\frac{\gamma_{d}}{1-\Omega_{d}}=\omega_{d}(\Omega_{d})-(\gamma-1)+\frac{\gamma_{d}}{1-\Omega_{d}},\\
&&f_{3}=f_{n}+\frac{\gamma_{tot}}{\Omega_{d}(1-\Omega_{d})}=\omega_{d}(\Omega_{d})-(\gamma-1)+\frac{\gamma_{tot}}{\Omega_{d}(1-\Omega_{d})},
 \end{eqnarray}

In conclusion, the fractional energies evolve as
\begin{eqnarray}
\label{mm1}
&& \Omega_{m}'= 3f_{j}\Omega_{m}\Omega_{d},\\
\label{dd1}
 && \Omega_{d}'= -3f_{j}\Omega_{d}\Omega_{m}.
 \end{eqnarray}
When $j$ takes $n$, it is the non-interacting case, and when $j$ is
$1,2,3$, it is the different interacting cases presented in
Eqs.(\ref{case1}), (\ref{case2}) and (\ref{case3}). This is an
autonomous system with  Eq.(\ref{constraint}) as its constraint.

\section{The Critical Points for the General Form}\label{sec4}
 In the following, we firstly obtain the critical points of the autonomous system by
imposing the conditions $\Omega_{m}'=\Omega_{d}'=0$. Obviously, the
critical points  should satisfy the Friedmann constraint
$\Omega_{m}+\Omega_{d}=1$.  Then, we will discuss the existence and
stability of these critical points. An attractor is one of the
stable critical points of the autonomous system.

%==================== table 2 ====================
\begin{table*}[t]
\begin{center}
\label{tab2}
 \caption[crit]{ Four main critical points.}
\begin{tabular}{|c|c|c|c|c|c|}

    \hline\hline
Label&Physical  meaning&$(\Omega_{m},\Omega_{d})$&$ f_{j}$&$(\lambda_{1},\lambda_{2})$ &$\omega_{tot}$\\
\hline
M&Matter dominated&$(1,0)$&$f_{j}$&$(0,-3f_{j})$& $\omega_{tot}=\omega_{m}$\\
$F_{M}$&Matter dominated&$(\Omega_{m},\Omega_{d})$&$f_{j}=0$,
&$(0,-3f_{dj}'\Omega_{d}\Omega_{m})$ &
  $\omega_{tot}=\frac{\Omega_{m}\omega_{m}/\Omega_{d}+\omega_{d}}{1+\Omega_{m}/\Omega_{d}}$\\
 D&DE dominated& $(0,1)$&$f_{j}$&$(0,3f_{j}) $&$\omega_{tot}=\omega_{d}$\\
$F_{D}$&DE
 dominated&$(\Omega_{m},\Omega_{d})$&$f_{j}=0$,&$(0,-3f_{dj}'
\Omega_{d}\Omega_{m})$&$\omega_{tot}=\frac{\Omega_{m}\omega_{m}/\Omega_{d}+\omega_{d}}{1+\Omega_{m}/\Omega_{d}}$\\
 \hline
 \hline
 \end{tabular}
\end{center}
\end{table*}

Based on the history of the universe and Eqs.(\ref{mm1}),
(\ref{dd1}), we can get that the critical points in Tab.2. Critical
point $M$ is the matter dominated phase with $\Omega_{m}=1$. And
Critical point $D$ is the dark energy dominated phase with
$\Omega_{d}=1$. If $f_{j}\propto\frac{1}{\Omega_{m}}$ or
$f_{j}\propto\frac{1}{\Omega_{d}}$, these two fixed point may not
exist. Besides the above  two fixed points, there are other
solutions with $f_{j}=0$. As the universe goes through
 the matter dominated phase and the
dark energy dominated phase, we  consider the critical points  with
$\omega_{tot}=0$ or $\omega_{tot}=-1$, which is matter dominated
phase $F_{M}$ or dark energy dominated phase $F_{D}$.

If we substitute linear perturbations about the critical point
$(\Omega_{m},\Omega_{d})$ into the dynamical system Eqs. (\ref{mm1})
 and (\ref{dd1}) and linearize them,
we can get that
 \begin{eqnarray}
 &&\delta\Omega_{m}'=
3f(\Omega_{d})\Omega_{d}\delta\Omega_{m}+3\left(f(\Omega_{d})\Omega_{m}+\frac{d
f(\Omega_{d})}{d \Omega_{d}}\Omega_{m}\Omega_{d}\right)\delta\Omega_{d},\\
 &&\delta\Omega_{d}'=
-3f(\Omega_{d})\Omega_{d}\delta\Omega_{m}-3\left(f(\Omega_{d})\Omega_{m}+\frac{d
f(\Omega_{d})}{d
\Omega_{d}}\Omega_{m}\Omega_{d})\right)\delta\Omega_{d}.
 \end{eqnarray}
The two eigenvalues of the coefficient matrix of the above equations
determine the stability of the corresponding critical point, which
yield two eigenvalues
\begin{eqnarray}
 &&\lambda_{1}=0,\\
 &&\lambda_{2}=3f(2\Omega_{d}-1)-3f'_{d}\Omega_{d}(1-\Omega_{d}).
 \end{eqnarray}
Here $f'_{d}=d f/d \Omega_{d}$. When $\lambda_{2}$ is positive, the
corresponding critical point is an unstable node. ``Unstable" means
that the phase will not stay in the phase for long, and eventually
it will evolve to other phases. When $\lambda_{2}$ is negative, the
corresponding critical point is a stable node and the phase will
last long. We list these critical points in Tab.2 as well.

With regard to the history of the universe, we expect that the
matter dominated phase should be unstable; otherwise, the universe
will not enter the dark energy dominated phase. The stability of the
matter dominated phase will determine whether the model coincides
with the history of the universe or not. If the matter dominated
phase ($M$ or $F_{M}$ ) is stable, we will see that there is no way
for the universe to get into a dark energy dominated phase. The
stability of the dark energy dominated phase ($D$ or $F_{D}$) will
determine the fate of the universe. If it is stable, the dark energy
phase is an attractor that the universe will be dark energy
dominated in the future. With a preceding unstable matter-dominated
phase, the universe will enter the phase at last. If the critical
points are unstable, then, even through the EoS parameter might be
smaller than $-1$, the universe would escape the big rip
singularity.

\section{The Interacting Terms}
As we discussed,  the four critical points in Tab.2 may not all
exist, especially in the interaction cases.

With interacting term one, we will not have the critical point $M$.
With the interacting term two, we will not have the critical point
$D$.
 With the
interacting term three, we will not have both the critical points
$M$ and $D$. When $f_{j}=0$, from Eqs.(\ref{wt}) and (\ref{fj}), we
can get that
 \be
 \omega_{tot}=\omega_{de}=\omega_{me}.
 \ee
 There are two scaling solutions. That the solutions may appear in the
 non-interacting and interacting cases or not depends on the exact form of
 $\omega_{d}$.

At the critical points $F_{M}$, we assume $\omega_{tot}=0$ for a
matter dominated phase, then we can get that
 \begin{eqnarray}
&&\omega_{tot}=\gamma-1-\gamma_{m}=0,\\
&&\omega_{tot}=\gamma-1-\gamma_{d}\frac{\Omega_{d}}{1-\Omega_{d}}=0,\\
&&\omega_{tot}=\gamma-1-\frac{\gamma_{tot}}{1-\Omega_{d}}=0.
 \end{eqnarray}
In the interacting term one case,  $\gamma-1=\gamma_{m}$. In the
interacting term two case,
$\Omega_{d}=(\gamma-1)/(\gamma_{d}+\gamma-1))$ and in the
interacting term three case, $\Omega_{d}=1-\gamma_{tot}/(\gamma-1)$.

And, at the critical points $F_{D}$, as $\omega_{tot}=-1$,  we can
get
 \begin{eqnarray}
&&\omega_{tot}=\gamma-1-\gamma_{m}=-1,\\
&&\omega_{tot}=\gamma-1-\gamma_{d}\frac{\Omega_{d}}{1-\Omega_{d}}=-1,\\
&&\omega_{tot}=\gamma-1-\frac{\gamma_{tot}}{1-\Omega_{d}}=-1.
 \end{eqnarray}
 In the interacting term one case,  $\gamma=\gamma_{m}$. In the interacting term two case,
$\Omega_{d}=\gamma/(\gamma_{d}+\gamma))$ and in the interacting term
three case, $\Omega_{d}=1-\gamma_{tot}/\gamma$. Because
$0\leq\Omega_{d}\leq1$, the interaction parameter should  satisfy
 $\gamma_{d}>0$ and $\gamma_{tot}>0$.

If one of the critical points $F_{M}$ or $F_{D}$ exists, it can be
treated as a scaling solution which  is helpful to solve the
coincidence problem. And the above discussions show that the
phase-analysis method is effective to describe the dark energy
models even with interactions  included. In the following, we will
give exact examples.

\section{The Examples}\label{sec5}
In this section, we list five examples for three purposes, to choose
 models which  coincide with the history of the universe,  to consider the fate of the universe, and the last is to
consider the effects of the interactions. The five dark energy
examples are in Tab.1: the cosmological constant model, the
holographic dark energy model, the agegraphic dark energy model, the
modified holographic dark energy model and the Ricci scalar
holographic dark energy model.

\subsection{Cosmological Constant Model}
First, we investigate  the cosmological constant case with
$L=constant$. In the non-interacting case, we can see that
 \begin{eqnarray}
 &&\omega_{d}^{c}=-1,
 \,\,\,\omega_{m}=\gamma-1,\\
 &&f_{n}^{c}=-\gamma,
 \end{eqnarray}
 where the superscript  $c$ means the cosmological constant model.
If we put the above equations into the evolution of the energy
density we can get that
\begin{eqnarray}
\label{omc1}
&& \Omega_{m}'= -3\gamma\Omega_{m}\Omega_{d},\\
\label{odc1}
 && \Omega_{d}'=3\gamma\Omega_{d}\Omega_{m}.
 \end{eqnarray}
 In
Tab.3, we list the main results related to the critical points. The
non-interacting cosmological constant model gives out a unstable
matter dominated phase, and a dark energy attractor as well.

%==================== table c0 ====================
\begin{table*}[t]
\begin{center}
\label{tabc0}
 \caption[crit]{ The properties of the
critical points in the cosmological constant model.}
\begin{tabular}{|c|c|c|c|c|c|}

    \hline\hline
& $(\Omega_{m},\Omega_{d})$& $(\lambda_{1},\lambda_{2})$  & existence& stability&$\omega_{tot}$\\
\hline
 M&(1,0)&$(0,3\gamma)$&always & unstable &
 $\gamma-1$\\
 D&(0,1)&$(0,-3\gamma)$& always
 &stable&$-1$\\
\hline
 \hline
 \end{tabular}
\end{center}
\end{table*}

In the interacting term cases, we replace the EoS parameter
$\omega_{m}$ and $\omega_{d}$ with the effective EoS parameter
$\omega_{me}$ and $\omega_{de}$. In the interaction one case,
 \begin{eqnarray}
&&f_{1}^{c}=\omega_{de1}-\omega_{me1}=-\gamma+\frac{\gamma_{m}}{\Omega_{d}}.\\
&&f_{d1}^{c'}=-\frac{\gamma_{m}}{\Omega_{d}^{2}}.
 \end{eqnarray}
 In the interacting term two,
\begin{eqnarray}
&&f_{2}^{c}=-\gamma+\frac{\gamma_{d}}{(1-\Omega_{d})},\\
&&f_{d2}^{c'}=\frac{\gamma_{d}}{(1-\Omega_{d})^{2}}.
\end{eqnarray}
In the interacting term three,
\begin{eqnarray}
&&f_{3}^{c}=-\gamma+\frac{\gamma_{tot}}{\Omega_{d}(1-\Omega_{d})},\\
&&f_{d3}^{c'}=\gamma_{tot}(-\frac{1}{\Omega_{d}^{2}}+\frac{1}{(1-\Omega_{d})^{2}}).
\end{eqnarray}
As are listed in Tabs.4, 5 and 6,  we  can always get the unstable
matter dominated
 phase and the stable dark energy dominated phase. We can see that
 the critical points require $\gamma_{m}>0$, $\gamma_{d}>0$ and
 $\gamma_{tot}>0$.  The
interactions bring further constraint to the parameter $\gamma$.

 %==================== tablec1====================
\begin{table*}[t]
\begin{center}
\label{tabc1}
 \caption[crit]{The properties of the
critical points in the cosmological constant model with interacting
term one.}
\begin{tabular}{|c|c|c|c|c|c|}
    \hline\hline
& $(\Omega_{m},\Omega_{d})$& $(\lambda_{1},\lambda_{2})$  & existence& stability&$\omega_{tot}$\\
\hline
$F_{M}$&$(1-\frac{\gamma_{m}}{\gamma},\frac{\gamma_{m}}{\gamma})$&$0,3\frac{\gamma_{m}\Omega_{m}}{\Omega_{d}})$&$0\leq\gamma_{m}\leq\gamma$&unstable&
 $0$\\
 D&(0,1)&$(0,-3(\gamma-\gamma_{m}))$& always
 &$\gamma_{m}<\gamma$, stable&$-1$\\
\hline
 \hline
 \end{tabular}
\end{center}
\end{table*}

%==================== table c2 ====================
\begin{table*}[t]
\begin{center}
\label{tabc2}
 \caption[crit]{ The properties of the
critical points in  the cosmological constant model with interacting
term two.}
\begin{tabular}{|c|c|c|c|c|c|}

    \hline\hline
& $(\Omega_{m},\Omega_{d})$& $(\lambda_{1},\lambda_{2})$  & existence& stability&$\omega_{tot}$\\
\hline
 M&(1,0)&$(0,3\gamma$& always
 &unstable&$\gamma-1$\\
 $F_{D}$&$(\frac{\gamma_{d}}{\gamma},1-\frac{\gamma_{d}}{\gamma})$&$(0,-3\frac{\gamma_{d}\Omega_{d}}{\Omega_{m}})$&
 $0\leq\gamma_{d}\leq\gamma $& stable&
 $-1$\\
\hline
 \hline
 \end{tabular}
\end{center}
\end{table*}

%==================== table c3 ====================
\begin{table*}[t]
\begin{center}
\label{tabc3}
 \caption[crit]{ The properties of the
critical points in  the cosmological constant model with interacting
term three. For conciseness, we define a new variable
$k=\sqrt{-\frac{\gamma_{tot}}{\gamma}+\frac{1}{4}}$.}
\begin{tabular}{|c|c|c|c|c|c|}
    \hline\hline
& $(\Omega_{m},\Omega_{d})$& $(\lambda_{1},\lambda_{2})$  & existence& stability&$\omega_{tot}$\\
\hline
$F_{M}$&$(\frac{1}{2}+k),\frac{1}{2}-k)$&$(0,-3f_{d3}^{h'}\Omega_{m}\Omega_{d})$&$0\leq\frac{\gamma_{tot}}{\gamma}\leq\frac{1}{4}$&$f_{d3}^{h'}<0$, unstable&$0$\\
 $F_D$&$(\frac{1}{2}-k),\frac{1}{2}+k)$&$(0,-3f_{d3}^{h'}\Omega_{m}\Omega_{d})$&$0\leq\frac{\gamma_{tot}}{\gamma}\leq\frac{1}{4}$&$f_{d3}^{h'}>0$, stable&$-1$\\
\hline
 \hline
 \end{tabular}
\end{center}
\end{table*}

\subsection{Holographic Dark Energy}
The successive holographic dark energy model uses the future event
horizon of the universe as the IR cutoff, which leads to
\begin{eqnarray}
&&L=a\int_{a}^{\infty}\frac{dt'}{a(t)}\\
 &&\rho_{d}^{h}=3c^{2}m_{pl}^{2}/L^{2},
\end{eqnarray}
where the superscript  $h$ means the holographic dark energy model.

In the non-interacting term,
 \begin{eqnarray}
 &&\omega_{d}^{h}=\frac{-1}{3}(1+\frac{2}{c}\sqrt{\Omega_{d}}),
 \,\,\,\omega_{m}=\gamma-1,\\
 &&f_{n}^{h}=\omega_{d}-\omega_{m}=\frac{-1}{3}(1+\frac{2}{c}\sqrt{\Omega_{d}})-(\gamma-1).
 \end{eqnarray}

If we put the above equations into the evolution of the energy
density, we can list the main results related to the critical points
in Tab.7. In the non-interaction case, if $\gamma>2/3$, the
matter-dominated phase is unstable, in the meanwhile the dark energy
dominated phase is also unstable. For example, assuming $\gamma=1$,
$c>-1$ will make the unstable case, which means we can get the
big-rip fate in holographic dark energy and the realization of the
dark energy dominated phase  hardly depends on the initial
conditions.

%==================== table h0 ====================
\begin{table*}
\begin{center}
 \caption[crit]{ The properties of the
critical points in holographic dark energy model.}
\begin{tabular}{|c|c|c|c|c|c|}

    \hline\hline
Label& $(\Omega_{m},\Omega_{d})$& $(\lambda_{1},\lambda_{2})$  & existence& stability&$\omega_{tot}$\\
\hline
 M&(1,0)&$(0,\gamma-\frac{2}{3})$&always & $\gamma>\frac{2}{3}$,unstable &
 $\gamma-1$\\
 D&(0,1)&$(0,2(1-\frac{1}{c})-3\gamma)$& always
 &$c>\frac{2}{2-3\gamma}$,unstable&$-\frac{1}{3}(1+\frac{1}{c})$\\
\hline
 \hline
 \end{tabular}
\end{center}
\end{table*}

With  the interacting term one,
 \begin{eqnarray}
&&f_{1}^{h}=\omega_{de1}-\omega_{me1}=\frac{-1}{3}(1+\frac{2}{c}\sqrt{\Omega_{d}})-(\gamma-1)+\frac{\gamma_{m}}{\Omega_{d}},\\
&&f_{d1}^{h'}=\frac{-1}{3c\sqrt{\Omega_{d}}}-\frac{\gamma_{m}}{\Omega_{d}^{2}}.
 \end{eqnarray}
We can get the critical point $D$, and the existence of $\gamma_{m}$
extends the parameter regime of $c$. There is no $M$ point because
of the interaction. However, there is a new matter dominated phase
$F_{M}$. To get such a phase, we require $\omega_{tot}=0$ and
$f_{1}^{h}=0$, so we can find  when the total universe behaves like
matter-dominated. As the matter dominated, we need $\Omega_{d}$ to
be much smaller than $\Omega_{m}$. Specifically speaking ,we can
assume $\Omega_{d}\ll c^{2}/2$,
$\Omega\approx\frac{\gamma_{m}}{-2/3+\gamma}$. $\gamma_{m}>0$ could
lead to the unstable matter dominated phase.
 And we set
$\gamma_{m}=\gamma-1$, which means $\omega_{tot}=0$.

%==================== tableh1====================
\begin{table*}[t]
\begin{center}
\label{tabh1}
 \caption[crit]{ The properties of the
critical points in holographic dark energy model with interacting
term one. For conciseness,
$\Omega_{d}=\frac{\gamma_{m}}{-2/3+\gamma}$ for the $F_{M}$ phase is
presented in the caption.}
\begin{tabular}{|c|c|c|c|c|c|}

    \hline\hline
Label& $(\Omega_{m},\Omega_{d})$& $(\lambda_{1},\lambda_{2})$  & existence& stability&$\omega_{tot}$\\
\hline
$F_{M}$&$(1-\Omega_{d},\Omega_{d})$&$0,3\Omega_{m}(\frac{\sqrt{\Omega_{d}}}{c}+
 \frac{\gamma_{m}}{\Omega_{d}})$&$\frac{\gamma-\frac{2}{3}}{\gamma-1}\ll\frac{c^{2}}{2}$&$\gamma_{m}>0$,unstable&
 $0$\\
 D&(0,1)&$(0,2(1-\frac{1}{c})-3\gamma+3\gamma_{m})$& always
 &$c>\frac{2}{2-3\gamma+3\gamma_{m}}$,unstable&$\frac{-1-2c}{3c}$\\
\hline
 \hline
 \end{tabular}
\end{center}
\end{table*}

%==================== table h2 ====================
\begin{table*}[t]
\begin{center}
\label{tabh2}
 \caption[crit]{ The properties of the
critical points in holographic dark energy model with interacting
term two.}
\begin{tabular}{|c|c|c|c|c|c|}

    \hline\hline
Label& $(\Omega_{m},\Omega_{d})$& $(\lambda_{1},\lambda_{2})$  & existence& stability&$\omega_{tot}$\\
\hline
 M&(1,0)&$(0,-3(\frac{2}{3}-\gamma-\gamma_{d}))$& always
 &$\gamma>\frac{2}{3}+\gamma_{d}$,unstable&$\gamma_{m}+\gamma-1$\\
 $F_{D}$&$(\frac{\gamma_{d}}{\gamma_{d}+\gamma},\frac{\gamma}{\gamma_{d}+\gamma})$&$(0,-3(\frac{\gamma_{d}\Omega_{d}}{\Omega_{m}}-\frac{\Omega_{m}\sqrt{\Omega_{d}}}{3c}))$&
 $\gamma_{d}>0 $&$\frac{\gamma_{d}^{2}}{9\gamma (\gamma_{d}+\gamma)^{3}}<c^{2}$, stable&
 $-1$\\
\hline
 \hline
 \end{tabular}
\end{center}
\end{table*}

In the interacting term two case,
\begin{eqnarray}
&&f_{2}^{h}=\frac{-1}{3}(1+\frac{2}{c}\sqrt{\Omega_{d}})-(\gamma-1)+\frac{\gamma_{d}}{(1-\Omega_{d})},\\
\label{h2p}
&&f_{d2}^{h'}=\frac{-1}{3c\sqrt{\Omega_{d}}}+\frac{\gamma_{d}}{(1-\Omega_{d})^{2}}.
\end{eqnarray}
The critical point $M$ shows that if $\gamma>2/3+\gamma_{d}$, the
matter dominated phase is unstable. And the $F_{D}$ critical point
shows that we can get a dark energy  attractor in holographic model
when the first term in the Eq.(\ref{h2p}) is much smaller than the
second term.

With the interacting term three,
\begin{eqnarray}
&&f_{3}^{h}=\frac{-1}{3}(1+\frac{2}{c}\sqrt{\Omega_{d}})-(\gamma-1)+\frac{\gamma_{tot}}{\Omega_{d}(1-\Omega_{d})},\\
\label{h3p}
&&f_{d3}^{h'}=-\frac{1}{3c\sqrt{\Omega_{d}}}-\frac{\gamma_{tot}}{\Omega_{d}^{2}}+\frac{\gamma_{tot}}{(1-\Omega_{d})^{2}}.
\end{eqnarray}
 Firstly, as is noted in Tab.10 , the existence of a critical point $F_{D}$ needs $\gamma_{tot}>0$. When
$\Omega_{d}$ is large, the absolute value of the
 third term in Eq.(\ref{h3p}) will be larger than the first  two terms,
$f_{d3}^{h'}>0$, and we get the stable dark energy dominated phase.
Instead, if $\Omega_{d}$ is small,   $f_{d3}^{h'}<0$, we will get
the unstable matter dominated phase.

%==================== table h3 ====================
\begin{table*}[t]
\begin{center}
\label{tabh3}
 \caption[crit]{ The properties of the
critical points in holographic dark energy model with interacting
term three.}
\begin{tabular}{|c|c|c|c|c|c|}
    \hline\hline
Label& $(\Omega_{m},\Omega_{d})$& $(\lambda_{1},\lambda_{2})$  & existence& stability&$\omega_{tot}$\\
\hline
$F_{M}$&$(\frac{\gamma_{tot}}{\gamma-1},1-\frac{\gamma_{tot}}{\gamma-1})$&$(0,-3f_{d3}^{h'}\Omega_{m}\Omega_{d})$&$0\leq\frac{\gamma_{tot}}{\gamma-1}\leq1$&$f_{d3}^{h'}<0$, unstable&$0$\\
 $F_D$&$(\frac{\gamma_{tot}}{\gamma},1-\frac{\gamma_{tot}}{\gamma})$&$(0,-3f_{d3}^{h'}\Omega_{m}\Omega_{d})$&$0\leq\frac{\gamma_{tot}}{\gamma}\leq1$&$f_{d3}^{h'}>0$, stable&$-1$\\
\hline
 \hline
 \end{tabular}
\end{center}
\end{table*}
We can see that in holographic dark energy model, as we predicted,
we can get the matter dominated and the dark energy dominated phase.
 The  interacting term could play an important role to make the
dark energy dominated phase stable.

 \subsection{Agegraphic Dark Energy}\label{sub2}
Agegraphic dark energy model, whose cutoff is  the age the universe
by the  age of the universe, is proved to be consistent with the
evolution of universe,
 \begin{eqnarray}
 &&L=T=\int_{0}^{a}\frac{da}{Ha},\\
&&\rho_{d}^{a}=\frac{3n^{2}m_{pl}^{2}}{T^{2}},
  \end{eqnarray}
where the superscript  $a$ means the agegraphic dark energy model.

In the non-interaction case,
 \begin{eqnarray}
 &&\omega_{d}^{a}=-1+\frac{2\sqrt{\Omega_{d}}}{3n},\\
 &&f_{n}^{a}=\frac{2}{3n}\sqrt{\Omega_{d}}-\gamma,\\
 &&f_{dn}^{a'}=\frac{1}{3n\sqrt{\Omega_{d}}}.
 \end{eqnarray}
Putting these forms in the Eq.(\ref{mm1}) and (\ref{dd1}), we could
get the stability of the critical points which is listed in Tab.11.

%==================== table a0 ====================
\begin{table*}[t]
\begin{center}
\label{taba0}
 \caption[crit]{ The properties of the
critical points in agegraphic dark energy model.}
\begin{tabular}{|c|c|c|c|c|c|}

    \hline\hline
Label& $(\Omega_{m},\Omega_{d})$& $(\lambda_{1},\lambda_{2})$  & existence& stability&$\omega_{tot}$\\
\hline
 M&(1,0)&$(0,3\gamma)$&always & unstable &$\gamma-1$\\
 $F_{M}$&$(1-\frac{9n^{2}\gamma^{2}}{4},\frac{9n^{2}\gamma^{2}}{4})$&$(0,-\frac{3\gamma}{2}(1-\frac{9n^{2}\gamma^{2}}{4}))$&$n<\frac{2}{3\gamma}$&stable&$\gamma-1$\\
 D&(0,1)&$(0,\frac{2}{n}-3\gamma)$& always
 &$n>\frac{2}{3\gamma}$, stable&$-1+\frac{2}{3n}$\\
\hline
 \hline
 \end{tabular}
\end{center}
\end{table*}
In Tab.11, we can see that, there is an interesting critical point
$F_{M}$ which is an attractor, so it is impossible to  have the dark
energy dominated phase. There are three points. With  $n$ larger
than $2/3\gamma$,  it could get the dark energy dominated phase;
otherwise, there will  always be matter dominated. And to make the
universe accelerate, in the dark energy dominated phase
$\omega_{tot}=-1+2/3n<-1/3$, it requires $n>1$.

With the interacting case one,
\begin{eqnarray}
&&f_{1}^{a}=\frac{2}{3n}\sqrt{\Omega_{d}}-\gamma+\frac{\gamma_{m}}{\Omega_{d}},\\
&&f_{d1}^{a'}=\frac{\Omega_{d}^{3/2}-3n\gamma_{m}}{3n\Omega_{d}^{2}},
\end{eqnarray}
we can get the critical point $D$, and the existence of $\gamma_{m}$
extends the parameter regime of $n$. There is no $M$ point because
of the interaction. However, there is another matter dominated phase
$F_{M}$. To get such a phase, we require $\omega_{tot}=0$ and
$f_{1}^{a}=0$. As matter dominated, we could assume $\Omega_{d}$
much smaller than $\Omega_{m}$. Then we can get the unstable matter
dominated phase and the dark energy dominated phase.

%==================== table a1 ====================
\begin{table*}[t]
\begin{center}
\label{taba1}
 \caption[crit]{ The properties of the
critical points in agegraphic dark energy model  with interacting
case one.}
\begin{tabular}{|c|c|c|c|c|c|}

    \hline\hline
Label& $(\Omega_{m},\Omega_{d})$& $(\lambda_{1},\lambda_{2})$  & existence& stability&$\omega_{tot}$\\
\hline
$F_{M}$&$(1-\frac{\gamma_{m}}{\gamma},\frac{\gamma_{m}}{\gamma})$&$(0,\Omega_{m}(3\gamma-\sqrt{\frac{\gamma_{m}}{n^{2}\gamma}}))$
 &$\gamma_{m}=\gamma-1$&$\gamma_{m}<9\gamma^{3}n^{2}$,unstable&$0$\\
 D&(0,1)&$(0,\frac{2}{n}-3\gamma+3\gamma_{m})$& always
 &$n>\frac{2}{3\gamma-3\gamma_{m}}$,stable&$-1+\frac{2}{3n}$\\
\hline
 \hline
 \end{tabular}
\end{center}
\end{table*}

With the interacting term  two,
\begin{eqnarray}
&&f_{2}^{a}=-1+\frac{2}{3n}\sqrt{\Omega_{d}}-(\gamma-1)+\frac{\gamma_{d}}{(1-\Omega_{d})}=0,\\
&&f_{d2}^{a'}=\frac{(1-\Omega_{d})^{2}+3n\gamma_{d}\sqrt{\Omega_{d}}}{3n\sqrt{\Omega_{d}}(1-\Omega_{d})^{2}}.
\end{eqnarray}
In Tab.12, it only needs that $0<\gamma_{d}<\gamma$, and we can get
the unstable matter dominated phase and the stable dark energy
dominated phase.

%==================== table a2 ====================
\begin{table*}[t]
\begin{center}
\label{taba2}
 \caption[crit]{ The properties of the
critical points in agegraphic dark energy model in interaction two.}
\begin{tabular}{|c|c|c|c|c|c|}
    \hline\hline
Label& $(\Omega_{m},\Omega_{d})$& $(\lambda_{1},\lambda_{2})$  & existence& stability&$\omega_{tot}$\\
\hline M&(1,0)&$(0,-\gamma_{d}+\gamma)$& always
 &$\gamma_{d}<\gamma$,unstable&$\gamma-1$\\
 $F_{D}$&$(\frac{\gamma_{d}}{\gamma+\gamma_{d}},\frac{\gamma}{\gamma+\gamma_{d}})$&$(0,-3(\frac{\sqrt{\Omega_{d}}\Omega_{m}}{3n}+\frac{\gamma_{d}\Omega_{d}}{\Omega_{m}}))$&$\gamma_{d}>0$
 &stable&$-1$\\
\hline
 \hline
 \end{tabular}
\end{center}
\end{table*}

With the interacting term three,
\begin{eqnarray}
&&f_{3}^{a}=\frac{2}{3n}\sqrt{\Omega_{d}}-\gamma+\frac{\gamma_{tot}}{\Omega_{d}(1-\Omega_{d})},\\
\label{a3p}
&&f_{d3}^{a'}=\frac{1}{3n\sqrt{\Omega_{d}}}+\frac{\gamma_{tot}}{\Omega_{d}^{2}\Omega_{m}}-\frac{\gamma_{tot}}{\Omega_{d}\Omega_{m}^{2}}.
\end{eqnarray}
The existence of the critical point $F_{D}$ needs $\gamma_{tot}>0$.
When $\Omega_{tot}$ is large, the absolute value of the
 third term in Eq.(\ref{a3p}) will be larger than the first  two terms,
$f_{d3}^{a'}>0$, then we get the stable dark energy dominated phase.
Instead, if $\Omega_{d}$ is small,   $f_{d3}^{a'}<0$, then we will
get the unstable matter dominated phase.

%%==================== table a3 ====================
\begin{table*}[t]
\begin{center}
\label{taba3}
 \caption[crit]{ The properties of the
critical points in agegraphic dark energy model in interaction
three.}
\begin{tabular}{|c|c|c|c|c|c|}

    \hline\hline
Label& $(\Omega_{m},\Omega_{d})$& $(\lambda_{1},\lambda_{2})$  & existence& stability&$\omega_{tot}$\\
\hline
$F_{M}$&$(\frac{\gamma_{tot}}{\gamma-1},1-\frac{\gamma_{tot}}{\gamma-1})$&$(0,-3f_{d3}^{a'}\Omega_{m}\Omega_{d})$&$0\leq\frac{\gamma_{tot}}{\gamma-1}\leq1$&$f_{d3}^{a'}<0$, unstable&$0$\\
 $F_D$&$(\frac{\gamma_{tot}}{\gamma},1-\frac{\gamma_{tot}}{\gamma})$&$(0,-3f_{d3}^{a'}\Omega_{m}\Omega_{d})$&$0\leq\frac{\gamma_{tot}}{\gamma}\leq1$&$f_{d3}^{a'}>0$, stable&$-1$\\
\hline
 \hline
 \end{tabular}
\end{center}
\end{table*}

 In  conclusion, the interaction is very helpful to extend
 the parameter phase and then alleviate the coincidence problem.

 \subsection{Modified Hologrphic Dark Energy}
 %==================== table m0 ====================
\begin{table*}[t]
\begin{center}
\label{tabm0}
 \caption[crit]{ The properties of the
critical points in Modified Holographic dark energy model.}
\begin{tabular}{|c|c|c|c|c|c|}

    \hline\hline
Label& $(\Omega_{m},\Omega_{d})$& $(\lambda_{1},\lambda_{2})$  & existence& stability&$\omega_{tot}$\\
\hline
 M&(1,0)&$(0,3-\frac{3}{2}\alpha\gamma)$&always &$\alpha<\frac{2}{\gamma}$, unstable &$\gamma-1$\\
 D&(0,1)&$(0,-3)$& always&stable&$-1$\\
\hline
 \hline
 \end{tabular}
\end{center}
\end{table*}

%==================== table m1 ====================
\begin{table*}[t]
\begin{center}
\label{tabm1}
 \caption[crit]{ The properties of the
critical points in Modified Holographic dark energy model in
interacting case one. For conciseness,
$\Omega_{d}=\frac{2\gamma_{m}}{2\gamma-\alpha\gamma+\alpha\gamma_{m}}$
for the $F_{M}$ phase is presented in the caption. }
\begin{tabular}{|c|c|c|c|c|c|}

    \hline\hline
Label& $(\Omega_{m},\Omega_{d})$& $(\lambda_{1},\lambda_{2})$  &
existence& stability&$\omega_{tot}$\\
$F_{M}$&$(1-\Omega_{d},\Omega_{d})$
 &$(0,-3f_{d1}^{m'}\Omega_{d}\Omega_{m})$&$0\leq\Omega_{d}\leq1$&$f_{d1}^{m'}<0$,unstable&$-\gamma_{m}+\gamma-1$\\
D&(0,1)&$(0,3(\gamma_{m}-\gamma))$&always&$\gamma_{m}<\gamma$,
stable &
$-1$\\
\hline
 \hline
 \end{tabular}
\end{center}
\end{table*}

%==================== table m2 ====================
\begin{table*}[t]
\begin{center}
\label{tabm2}
 \caption[crit]{ The properties of the
critical points in Modified Holographic dark energy model in
interacting case two.}
\begin{tabular}{|c|c|c|c|c|c|}

    \hline\hline
Label& $(\Omega_{m},\Omega_{d})$& $(\lambda_{1},\lambda_{2})$  &
existence& stability&$\omega_{tot}$\\
M&(1,0)&$(0,-3(\frac{\alpha\gamma}{2}+\gamma_{d}-\gamma))$&always&$\gamma_{d}<\frac{(2-\alpha)\gamma}{2}$,
unstable &
$\gamma-1$\\
 $F_D$&$(\frac{\gamma_{d}}{\gamma_{d}+\gamma},\frac{\gamma}{\gamma_{d}+\gamma}))$&$(0,-3f_{d2}^{m'}\Omega_{d}\Omega_{m})$&$\gamma_{d}>0$&$f_{d2}^{m'}>0$, stable&$-1$ \\ \hline
 \hline
 \end{tabular}
\end{center}
\end{table*}

%==================== table m3 ====================
\begin{table*}[t]
\begin{center}
\label{tabm3}
 \caption[crit]{ The properties of the
critical points in modified holographic dark energy model in
interacting case three.}
\begin{tabular}{|c|c|c|c|c|c|}

    \hline\hline
& $(\Omega_{m},\Omega_{d})$& $(\lambda_{1},\lambda_{2})$  & existence& stability&$\omega_{tot}$\\
\hline
$F_{M}$&$(\frac{\gamma_{tot}}{\gamma-1},1-\frac{\gamma_{tot}}{\gamma-1})$&$(0,-3f_{d3}^{m'}\Omega_{m}\Omega_{d})$&$0\leq\frac{\gamma_{tot}}{\gamma-1}\leq1$&$f_{d3}^{m'}<0$, unstable&$0$\\
 $F_D$&$(\frac{\gamma_{tot}}{\gamma},1-\frac{\gamma_{tot}}{\gamma})$&$(0,-3f_{d3}^{m'}\Omega_{m}\Omega_{d})$&$0\leq\frac{\gamma_{tot}}{\gamma}\leq1$&$f_{d3}^{m'}>0$, stable&$-1$\\
\hline
 \hline
 \end{tabular}
\end{center}
\end{table*}
Furthermore, to realize a workable dark energy model, Gong use the
Hubble scale as the IR cut-off $L$, and the UV and IR connection is
modified by using the black hole mass $M$ in higher dimensions,
 \begin{eqnarray}
\label{uvir} && L^3\rho_d^{m}\sim M=\frac{(N-1)\Omega_{N-1}}{16\pi
G_N}L^{N-2},\\
&& \label{mholrd} \rho_d^{m}= \frac{d(N-1)\Omega_{N-1}}{16\pi
G_N}L^{N-5},
  \end{eqnarray}
where the superscript  $m$ means the modified holographic dark
energy model, $N$ is the number of spatial dimensions, $d$ is the
unknown constant related to the theoretical uncertainties, $G_{N}$
is the Newton constant and $\Omega_{N-1}$ is the volume of the black
hole.

In the non-interaction case,
 \begin{eqnarray}
&&\omega_{d}^{m}=-1+\frac{\alpha\gamma(1-\Omega_{d})}{2-\alpha\Omega_{d}},\omega_{m}=\gamma-1,\\
&&f_{n}^{m}=\frac{\alpha\gamma(1-\Omega_{d})}{2-\alpha\Omega_{d}}-\gamma,
  \end{eqnarray}
where $\alpha=5-N<2$.

In the non-interaction case, the critical point $M$ which presents
the matter dominated phase would be unstable only when
$\alpha<\frac{2}{\gamma}$. If $\gamma=1$, it means that, only for
the spatial dimension $N>3$, the universe could escape from the
matter dominated phase. When the spatial dimension is $N=3$, there
exists a matter-dominated phase $F_{M}$ again. We should calculate
to the second order perturbation to justify its stability.
Nevertheless, the dark energy dominated phase $D$ is an attractor.

With  the interaction term one,
\begin{eqnarray}
&&f_{1}^{m}=\frac{\alpha\gamma(1-\Omega_{d})}{2-\alpha\Omega_{d}}-\gamma+\frac{\gamma_{m}}{\Omega_{d}}=\frac{(\alpha-2)\gamma}{2-\alpha\Omega_{d}}+\frac{\gamma_{m}}{\Omega_{d}},\\
\label{mf1p}
&&f_{d1}^{m'}=-\frac{\gamma_{m}}{\Omega_{d}^{2}}-\frac{\alpha\gamma(2-\alpha)}{(2-\alpha\Omega_{d})^{2}}.
\end{eqnarray}
By adding the interacting  one, we will still get two phases. The
matter dominated phase $F_M$ is unstable  when $f_{d1}^{m'}<0$. We
need that $0\leq\alpha\leq2$ and $\gamma_{m}>0$. Furthermore,  with
$\gamma_{m}<\gamma $, the dark energy dominated phase is an
attractor.

With  the interaction term two,
\begin{eqnarray}
&&f_{2}^{m}=\frac{(\alpha-2)\gamma}{2-\alpha\Omega_{d}}+\frac{\gamma_{d}}{1-\Omega_{d}},\\
\label{m2p}
&&f_{d2}^{m'}=\frac{\gamma_{d}}{(1-\Omega_{d})^{2}}+\frac{\alpha\gamma(\alpha-2)}{(2-\alpha\Omega_{d})^{2}}.
\end{eqnarray}
From Tab.17, the instability of the matter dominated phase and the
existence of the dark energy dominated phase $F_{D}$ require that
$0<\gamma_{d}<(2-\alpha)\gamma/2$, and it means $\alpha<2$. Then for
a stable dark energy phase, we need that the first term in
Eq.(\ref{m2p}) is larger than the second term.

With  the interacting case three,
\begin{eqnarray}
&&f_{3}^{m}=\frac{(\alpha-2)\gamma}{2-\alpha\Omega_{d}}+\frac{\gamma_{tot}}{\Omega_{d}(1-\Omega_{d})},\\
\label{m3p}
&&f_{d3}^{m'}=\frac{\gamma_{tot}}{\Omega_{d}^{2}\Omega_{m}}+\frac{\alpha\gamma(\alpha-2)}{(2-\alpha\Omega_{d})^{2}}-\frac{\gamma_{tot}}{\Omega_{d}\Omega_{m}^{2}}
\end{eqnarray}

The existence of the critical point $F_{D}$ needs $\gamma_{d}>0$. If
$\alpha<2$, when $\Omega_{d}$ is large, the absolute value of the
 third term in Eq.(\ref{m3p}) will be larger than the first  two
 terms, and
$f_{d3}^{m'}>0$, we get the stable dark energy dominated phase.
Instead, if $\Omega_{d}$ is small,   $f_{d3}^{m'}<0$, we will get
the unstable matter dominated phase.  If $\alpha>2$, when
$\Omega_{d}$ is large, the absolute value of the
 last two terms in Eq.(\ref{m3p}) will be larger than the first term,
$f_{d3}^{m'}>0$, and we get the stable dark energy dominated phase.
Instead, if $\Omega_{d}$ is small,   $f_{d3}^{m'}<0$, we will get
the unstable matter dominated phase.

In  conclusion, the interaction one and three cases alleviate the
constraint on the spatial dimensions.

\subsection{Ricci  Dark Energy}
Recently, the average radius of the Ricci scalar curvature has been
chosen as the IR cutoff in  Ricci DE model:
     \begin{eqnarray}
     L=R=6(\dot{H}+H^{2})
   \rho_{d}^{r}=3\beta^{2}m_{pl}^{2}(\dot{H}+H^{2}),
   \end{eqnarray}
   where the superscript  $m$ means the modified holographic dark
energy model and $\beta$ is the model parameter. We can get
  \begin{eqnarray}
  &&\omega_{d}^{r}=-\frac{2}{3\beta^{2}}+\frac{1}{3\Omega_{d}}-\frac{\Omega_{m}}{\Omega_{d}}(\gamma-1),\,\,\omega_{m}=\gamma-1,\\
  &&f_{n}^{r}=-\frac{2}{3\beta^{2}}+\frac{1}{3\Omega_{d}}-\frac{1}{\Omega_{d}}(\gamma-1).\\
  &&f_{dn}^{r'}=(\gamma-\frac{4}{3})\frac{1}{\Omega_{d}^{2}}.
  \end{eqnarray}

In the non-interaction case, for the form of the $\omega_{d}^{r}$,
we could not get the $M$ critical point, but an $F_{M}$ critical
point instead. From Tab.19,  we can see the two critical points
which could lead us to the stable dark energy dominated phase and
the unstable matter dominated phase.

%==================== table r0 ====================
\begin{table*}[t]
\begin{center}
\label{tabr0}
 \caption[crit]{ The properties of the
critical points in non-interaction Ricci dark energy case.  For
conciseness, $\Omega_{d}=\frac{\beta^{2}(4-3\gamma)}{2}$ for the
$F_{M}$ phase is presented in the caption.}
\begin{tabular}{|c|c|c|c|c|c|}

    \hline\hline
Label& $(\Omega_{m},\Omega_{d})$& $(\lambda_{1},\lambda_{2})$  & existence& stability&$\omega_{tot}$\\
\hline $F_{M}$&$(1-\Omega_{d},
\Omega_{d})$&$(0,-3f_{n}^{r'}\Omega_{d}\Omega_{m})$&$0\leq\frac{\beta^{2}(4-3\gamma)}{2}\leq1$&$\gamma<4/3$,
unstable&$\gamma-1$\\
 D&(0,1)&$(0,-\frac{2}{3\beta^{2}}+\frac{4}{3}-\gamma)$& always &$\frac{1}{\beta^{2}}>2-\frac{3}{2}\gamma$,stable&$\frac{\beta^{2}-2}{3\beta^{2}}$\\
\hline
 \hline
 \end{tabular}
\end{center}
\end{table*}

In  the interacting case one,
\begin{eqnarray}
&&f_{1}^{r}=-\frac{2}{3\beta^{2}}+\frac{1}{3\Omega_{d}}-\frac{1}{\Omega_{d}}(\gamma-1)+\frac{\gamma_{m}}{\Omega_{d}},\\
&&f_{d1}^{r'}=(-\gamma_{m}+\gamma-\frac{4}{3})\frac{1}{\Omega_{d}^{2}}.
\end{eqnarray}
$\beta^{2}<1/3$, the universe could accelerate. And with certain
constraint on the parameter $\beta$, we can get the expected
unstable matter dominated phase and a stable dark energy dominated
phase.
%==================== table r1 ====================
\begin{table*}[t]
\begin{center}
\label{tabr1}
 \caption[crit]{ The properties of the
critical points in  Ricci dark energy model with interacting case
one. For conciseness,
$\Omega_{d}=\frac{\beta^{2}(4-3\gamma+3\gamma_{m})}{2})$ in $F_{M}$
phase is presented  in the caption.}
\begin{tabular}{|c|c|c|c|c|c|}

    \hline\hline
Label& $(\Omega_{m},\Omega_{d})$& $(\lambda_{1},\lambda_{2})$  & existence& stability&$\omega_{tot}$\\
\hline
$F_{M}$&$1-\Omega_{d},\Omega_{d}$&$(0,-3f_{d1}^{r'}\Omega_{d}\Omega_{m})$&$0\leq\Omega_{d}\leq1$& $\gamma_{m}>\gamma-\frac{4}{3}$,unstable&$\gamma-1$\\
 D&(0,1)&$(0,\frac{-2+\beta^{2}(4-3\gamma+3\gamma_{m})}{3\beta^{2}}$& always &$\frac{1}{\beta^{2}}>\frac{4-3(\gamma-\gamma_{m})}{2}$,stable&$\frac{\beta^{2}-2}{3\beta^{2}}$\\
\hline
 \hline
 \end{tabular}
\end{center}
\end{table*}

In the interacting case two, we can see that
\begin{eqnarray}
&&f_{2}^{r}=-\frac{2}{3\beta^{2}}+\frac{1}{3\Omega_{d}}-\frac{1}{\Omega_{d}}(\gamma-1)+\frac{\gamma_{d}}{(1-\Omega_{d})}=0,\\
\label{r2p}
&&f_{d2}^{r'}=(\gamma-\frac{4}{3})\frac{1}{\Omega_{d}^{2}}+\frac{\gamma_{d}}{(1-\Omega_{d})^{2}}.
\end{eqnarray}
The existence of the critical point $F_{D}$ needs $\gamma_{d}>0$.
Assuming $\gamma<4/3$, when $\Omega_{d}$ is large, the absolute
value of the
 second term in Eq.(\ref{r2p}) will be larger than the first term,
$f_{d2}^{r'}>0$, and we get the stable dark energy dominated phase.
Instead, if $1-\Omega_{d}$ is large,   $f_{d2}^{r'}<0$, we will get
the unstable matter dominated phase.

%==================== table r2 ====================
\begin{table*}[t]
\begin{center}
\label{tabr2}
 \caption[crit]{ The properties of the
critical points in Ricci dark energy model in interacting case two.}
\begin{tabular}{|c|c|c|c|c|c|}

    \hline\hline
Label& $(\Omega_{m},\Omega_{d})$& $(\lambda_{1},\lambda_{2})$  & existence& stability&$\omega_{tot}$\\
\hline
$F_{M}$&$(\frac{\gamma_{d}}{\gamma_{d}+\gamma-1},\frac{\gamma-1}{\gamma_{d}+\gamma-1})$&$(0,-3f_{d2}^{r'}\Omega_{d}\Omega_{m})$&$0\leq\frac{\gamma_{d}}{\gamma_{d}+\gamma-1}\leq1$&$f_{d2}^{r'}<0$, unstable&$0$\\
 $F_D$&$(\frac{\gamma_{d}}{\gamma_{d}+\gamma},\frac{\gamma-1}{\gamma_{d}+\gamma})$&$(0,-3f_{d2}^{r'}\Omega_{d}\Omega_{m})$&$\gamma_{d}>0$&$f_{d2}^{r'}>0$, stable&$-1$\\
\hline
 \hline
 \end{tabular}
\end{center}
\end{table*}

With the interaction term three,
\begin{eqnarray}
&&f_{3}^{r}=-\frac{2}{3\beta^{2}}+\frac{1}{3\Omega_{d}}-\frac{1}{\Omega_{d}}(\gamma-1)+\frac{\gamma_{tot}}{\Omega_{d}(1-\Omega_{d})},\\
&&f_{d3}^{r'}=(\gamma-\gamma_{tot}-\frac{4}{3})\frac{1}{\Omega_{d}^{2}}+\frac{\gamma_{tot}}{\Omega_{m}^{2}}.
\end{eqnarray}
The existence of the critical point $F_{D}$ needs $\gamma_{tot}>0$.
Assuming $\gamma-\gamma_{tot}<4/3$, when $\Omega_{d}$ is large, the
absolute value of the
 second term in Eq.(\ref{r2p}) will be larger than the first term,
$f_{d2}^{r'}>0$, and we get the stable dark energy dominated phase.
Instead, if $1-\Omega_{d}$ is large,   $f_{d2}^{r'}<0$, we will get
the unstable matter dominated phase.

%==================== table r3 ====================
\begin{table*}[t]
\begin{center}
\label{tabr3}
 \caption[crit]{ The properties of the
critical points in Ricci dark energy  model in interacting case
three.}
\begin{tabular}{|c|c|c|c|c|c|}

    \hline\hline
Label& $(\Omega_{m},\Omega_{d})$& $(\lambda_{1},\lambda_{2})$  & existence& stability&$\omega_{tot}$\\
\hline
$F_{M}$&$(\frac{\gamma_{tot}}{\gamma-1},1-\frac{\gamma_{tot}}{\gamma-1})$&$(0,-3f_{d3}^{r'}\Omega_{m}\Omega_{d})$&$0\leq\frac{\gamma_{tot}}{\gamma-1}\leq1$&$f_{d3}^{r'}<0$, unstable&$0$\\
 $F_D$&$(\frac{\gamma_{tot}}{\gamma},1-\frac{\gamma_{tot}}{\gamma})$&$(0,-3f_{d3}^{r'}\Omega_{m}\Omega_{d})$&$0\leq\frac{\gamma_{tot}}{\gamma}\leq1$&$f_{d3}^{r'}>0$, stable&$-1$\\
\hline
 \hline
 \end{tabular}
\end{center}
\end{table*}

In  conclusion, in the Ricci dark energy model, we successfully get
the desired matter dominated and dark energy dominated phases. The
interactions extend the regime of the parameter and help to get some
scaling solutions.

\subsection{ Discussions}

However, in the above discussions, we conclude that the interactions
may have three-fold action: to make the unstable dark energy
dominated phase an attractor, to extend the regime of the parameter,
and  the last is to make out scaling solutions to alleviate the
coincidence problem.

 \section{Conclusion}
  In the paper, we suggest that we could construct the models in which
the equation of state is related to the fractional dark energy.
Then, considering the evolution of the universe,  we  analyze the
dynamical behavior in the holographic-like models. Since the
universe experiences the matter dominated phase and the dark energy
dominated phase, we focus on this two phases.

We have presented a phase-space analysis of the evolution for a
spatially flat FRW universe containing dark energy and matter. Based
on the established  history of the universe, we want to get an
unstable matter dominated phase and a stable dark energy dominated
phase. The unstable matter dominated phase is necessary. If the dark
energy dominated phase is unstable, it may depend on the initial
conditions that the universe first went through a matter dominated
phase, and then a dark energy dominated phase. But if it is an
attractor, we do not need to consider the initial conditions.

%%%%%%%%%%%%%%%%%%%%%%%%%%%%%%%%%%%%%%%%%%%%%%%%%%%%%%%%%%%%%%%%%%%%%%%%%%%%%%%%%%%%%%%%%%%%%%%%%%
%============================= acknowledgments ===================================

\section*{Acknowledgements}
We thank Prof. Rong-gen Cai, Prof.Yun-Gui Gong and Prof.Zong-Hong
Zhu for useful discussions. This work was supported by CQUPT under
Grant No.A2009-16, the Ministry of Science and Technology of China
national basic science Program (973 Project) under grant Nos. 2007CB
815401 and 2010CB833004, the National Natural Science Foundation of
China key project under grant Nos. 10533010 and 10935013, and the
Distinguished Young Scholar Grant 10825313, and the Natural Science
Foundation Project of CQ CSTC under grant No. 2009BA4050.

\appendix

%=============================references ===================================

\end{document}